\documentclass[letterpaper, 10 pt, conference]{ieeeconf}  

\IEEEoverridecommandlockouts                              

\usepackage{graphics} 
\usepackage{epsfig} 
\usepackage{mathptmx} 
\usepackage{times} 
\usepackage{amsmath} 
\usepackage{amssymb}  
\usepackage{graphicx}

\title{\LARGE \bf
A participatory design approach to using social robots for elderly care$^*$
}

\author{Barbara Sienkiewicz$^1$, Zuzanna Radosz-Knawa$^2$, Bipin Indurkhya$^3$ 
\thanks{*This research was supported in part by the National Science Centre, Poland, under the OPUS call in the Weave programme under the project number K/NCN/000142.}
\thanks{$^{1}$Barbara Sienkiewicz is with the Cognitive Science Department,
        Jagiellonian University in Kraków, Poland.
        {\tt\small barbara.sienkiewicz@student.uj.edu.pl}}
\thanks{$^{2}$Zuzanna Radosz-Knawa is with the Faculty of Health Sciences,
        Jagiellonian University Medical College, Kraków, Poland.
        {\tt\small zuzanna.radosz-knawa@uj.edu.pl}}
\thanks{$^{3}$Bipin Indurkhya is with the Cognitive Science Department,
        Jagiellonian University in Kraków, Poland.
        {\tt\small bipin.indurkhya@uj.edu.pl}}}

\begin{document}
\maketitle
\thispagestyle{empty}
\pagestyle{empty}

\begin{abstract}
We present our ongoing research on applying a participatory design approach to using social robots for elderly care. Our approach involves four different groups of stakeholders: the elderly, (non-professional) caregivers, medical professionals, and psychologists. We focus on card sorting and storyboarding techniques to elicit the concerns of the stakeholders towards deploying social robots for elderly care. This is followed by semi-structured interviews to assess their attitudes towards social robots individually. Then we are conducting two-stage workshops with different elderly groups to understand how to engage them with the technology and to identify the challenges in this task. 
\end{abstract}

\section{INTRODUCTION}
The demographics of the world population are rapidly shifting towards the elderly. According to the World Health Organization \cite{WHO2022}: 
``By 2030, 1 in 6 people in the world will be aged 60 years or over. At this time the share of the population aged 60 years and over will increase from 1 billion in 2020 to 1.4 billion. By 2050, the world’s population of people aged 60 years and older will double (2.1 billion). The number of persons aged 80 years or older is expected to triple between 2020 and 2050 to reach 426 million.'' 

This increased proportion of elderly population creates daunting challenges to the healthcare sector. As the prowess of digital technologies has been exponentially increasing in recent years, digital healthcare is emerging as a major arena in which such challenges are being tackled  \cite{sharma2020}. In our research, we are focusing on the technology of social robots to care for the elderly.

In recent years, there has been a great interest in deploying social robots to care for the elderly \cite{vercelli2018}. Social robots offer a number of advantages in this role including, 1) they are available all the time; 2) they do not get frustrated or lose their patience; 3) they can be customized for each individual, taking into account their particular physical and cognitive impairments; 4) persuasive techniques, such as nudge theory and metaphor-based interfaces, can be incorporated in their interactions; 5) using machine learning, they can adapt to the pattern of behaviour of each individual, and can employ optimal strategies to be effective; 6) with the proliferation of mobile platforms and 3-d printing technologies, the cost of designing and implementing robotic platforms is decreasing everyday; and so on.

However, many of these social robot platforms for the elderly take an approach where the technologists impose their ‘solutions’ on the target user group. Even when user studies are done, it is assumed that the evaluations based on some users will generalize to all users. Such patriarchal and universal approaches have recently been subjected to severe criticism \cite{rosner2018critical,shew2023against}.  Instead, a participatory design (also known as co-design) approach is recommended where the target user group is included in the design process, and many local solutions are developed that address specific needs of different user groups. A related approach is embodied design \cite{hook2018designing}: Unlike abstract design, the focus here is on physical body and movement. This is especially important for the elderly, as their bodies and their physical and cognitive abilities are hardly neurotypical.

In an embodied design approach, empathy and understanding of the diversity of user experiences becomes a key factor. This means that we need to account for not only individual differences in physical and cognitive skills but also the social, cultural, and emotional context in which users will use our artifacts and interventions. We strive to create solutions that not only meet users' specific needs, but also integrate with their everyday experiences and sense of well-being.
 
In the case of older people, this approach becomes particularly important due to the complexity of their needs and the diversity of life situations. Developing personalized solutions becomes not only a matter of functional effectiveness, but also a way to increase the sense of dignity and autonomy in old age. By using embodied design, we can therefore not only create practical tools, but also build relationships based on empathy and understanding, which is crucial in providing support for older people to maintain their quality of life.

In this paper, we outline our research on applying the participatory design approach to deploy social robots for the elderly care. The paper is organized as follows. In the next section, we provide a brief review of the existing approaches to using social robots for elderly care, emphasizing the need for a participatory design. 

In Sec. III we provide a short introduction to the participatory design approach, followed, in Sec. IV, by an introduction to the embodied design approach. Sec. V articulates our project goals. In Sec. VI, we review related research on applying a participatory design approach to deploying social robots with different user groups. Our participatory design approach is outlined in Secs. VII and VIII. Finally, conclusions and future research directions are discussed in Sec. IX.

\section{Existing approaches to using social robots for elderly care}

A number of comprehensive reviews of using social robots for elderly care have appeared in recent years. For example, Moro {\it et al.} \cite{moro2019social} reviewed how the dynamic social features of a robot, such as facial expressions and gestures, affect the interaction experience of cognitively impaired seniors during an assistive activity. They found that the participants generally preferred a human-like robot (as opposed to an avatar or just a device like a tablet). They also recommended a holistic and multimodal approach to designing healthcare platforms for the elderly, where social robots are complemented with signs and notices placed around their home, medication reminders, orientation devices, alarms, wearable devices, and so on.

A similar meta-review  \cite{gonzalez2021social} on the design of social robots and their interaction with patients found that most of these projects serve the elderly and children. Moreover, they helped these groups cope with issues such as dementia and autism spectrum disorder (ASD), and fight diseases like cancer and diabetes.

In another recent review, Weerarathna {\it et al.} \cite{weerarathna2023humanrobot} present evidence on how social robots can offer personalized, low-cost, home-based digital technology for both curative and preventive care. They predict that smooth human-robot collaboration will lead to more accessible and client-centered healthcare, though there are practical, ethical, and legal challenges.

Chita-Tegmark and Scheutz \cite{chita2021} argue
that socially assistive robots can help people with health conditions maintain positive social lives by supporting them in social interactions.
They articulate a framework of social mediation functions implemented in a social robot to meet the social needs of people with health conditions. Then they identify five roles that a social robot can play to assist the elderly: (a) changing how the person is perceived, (b) enhancing the social behavior of the person, (c) modifying the social behavior of others, (d) providing structure for interactions, and (e) changing how the person feels. 

A number of recent research papers have pointed out the need for an inclusive, participatory design approach for deploying social robots for elderly care. For example, Burema \cite{burema2022} argues that technology development is a socially constructed process that can reinforce problematic understandings of the elderly. She analyzed 96 academic publications to find stereotypical representation of the elderly: ``frail by default, independent by effort; silent and technologically illiterate; burdensome; and problematic for society.'' Burema argues against this kind of bias and suggests an inclusive design approach that incorporates the perspectives of the users seriously.

In a similar vein, Ostrowski {\it et al.} \cite{ostrowski2021long} present guidelines for co-designing a social robot for the elderly based on their 12-month project conducting such an activity. They proposed the following design principles: scenario specific exploration, long-term lived experiences, supporting multiple design activities, cultivating relationships, and employing divergent and convergent processes.

\section{Introduction to participatory design}

The idea of \textit{Participatory Design (PD)} evolved from a series of projects in Scandinavia during the 1970s and 1980s, where researchers and unions jointly engaged in efforts to explore ways of democratizing the introduction of technology and ensuring the quality of work and products for workers \cite{bjerknes1987computers}. 
Since then, PD has been widely adopted in numerous disciplines, each with its distinct set of goals. In some cases, the primary focus is on the quality of the final product, while in others, the goal is to prioritize the empowerment and active involvement of the participants. 

However, a general definition could be: "Participatory Design can be defined as a process of investigating, understanding, reflecting upon, establishing, developing, and supporting mutual learning between multiple participants in collective ‘reflection-inaction’. The participants typically undertake the two principal roles of users and designers where the designers strive to learn the realities of the users’ situation while the users strive to articulate their desired aims and learn appropriate technological means to
obtain them. \cite{simonsen2012routledge}.

Moreover, PD promotes the agenda of democracy, quality of life, and empowerment of people to play an active role in technological development \cite{dindler2021computational}. 

\section{Introduction to embodied design}

A concept related to PD is that of \textit{Embodied Design,} where one focuses on the bodily interactions with the artefact being designed \cite{hook2018designing}. An implication of the embodied design approach is that it brings in subjective perspectives, as bodies are different and act differently in the same environment. This is in contrast with the universal approach to design, where one assumes that a design based on some user studies will be effective for most, if not all, users. However, in the subjective approach of embodied design, we are very cautious about any generalization and, instead of seeking universal design solutions, we aim to find a multitude of designs, each adopted to the needs of a small user group. This concept was applied successfully in the past to design assistive tools for children with autism \cite{sampath2013}.

This approach is especially suited for the elderly, because we notice that in our elderly care center, different people have different physical and cognitive impairments, and therefore we need to assess these individual needs and develop personalized solutions.

Recent studies by Kamali et al \cite{el2023co} on embodiment design have shown the importance of involving different older adults in the design process. These studies reveal that older adults across Europe (Spain, the UK, Italy) exhibited varying preferences for voice assistants. This underscores the necessity for personalization and local solutions in designs intended for the elderly.

 Liu et al. \cite{liu2022research} emphasize the importance of designing with an awareness of the cognitive and physical interactions of the elderly with their environment. They suggested that product interaction design must not only accommodate the physiological changes that accompany aging, but also take advantage of these changes to improve the usability and accessibility of products for the elderly. The authors link physical behavior and cognitive processes, advocating for a design methodology that recognizes and satisfies the sensory, perceptual, and motor skills of the elderly population. 

Lee et al. \cite{lee2020embodied} investigate the impact of an embodied interactive system 'Move and Paint' on the engagement and behavior patterns of older adults in public spaces. Their research demonstrates how such technology not only facilitates increased participation among older adults in these settings, but also potentially improves their social interaction and physical activity.

\section{Related research on participatory design with social robots}

In recent years, many projects have taken the PD approach to designing social robots for different user groups. For example, Schaper and her colleagues have pioneered a number of participatory and embodied design techniques for working with children. In \cite{schaper2019fubimethod} they proposed a method based on six stages to participatory design interactive experiences based on Full-Body Interaction. In a later work, Schaper and Pares \cite{schaper2021codesign} propose novel participatory design techniques to induce embodied awareness of children, which were evaluated through a series of design workshops with a local theater school. They found that these theater-based techniques were quite effective in allowing children to incorporate the specific features of full-body interaction in their participatory design. Most recently, Schaper {\it et al.} \cite{schaper2023think4emcode} presented a framework for embodied participatory design methods and techniques for children, which highlights ten qualities: (1) embodied awareness, (2) reflective imaginary, (3) emergence, (4) embodied memory, (5) situated relationality, (6) contingency, (7) playful engagement, (8) play practice, (9) developmental scaffolds, and (10) social dialogue. 

Some participatory design techniques have also been used with the elderly. For example, Fraga Viera {\it et al.}\cite{fraga2020inclusive} investigated and designed prompts for social interactions between elderly people and their family members mediated through different types of technology. 

In another study, Tian {\it et al.} \cite{tian2021redesigning}, applied the PD methodology, to investigate how the perception and expectations of the participants change in response to robot failures.  
Meanwhile, Lee {\it et al.}\cite{lee2017steps} illustrate the positive impact of involving end-users, particularly older adults with depression, in the robot design process. Through participation in design workshops, these adults experienced enhanced self-confidence, underscoring the significant benefits of user inclusion in creating socially assistive technologies.

Finally, Rogers {\it et al.} \cite{rogers2022maximizing} 
provided an extensive review of
PD research on the deployment of social robots to assist older adults. In their review, they emphasized the use of mixed methods and the inclusion of multiple stakeholders throughout the design process. The approach we have taken in our project is heavily influenced by these findings.

\section{Project goals}

Taking into account our focus on the elderly, we are applying participatory design to deeply understand their lives and perspectives on social robots. Our aim is to collaboratively explore ways of deploying social robots that not only meet their needs but also familiarize them with technology, thereby reducing apprehension towards its use. By actively involving the elderly in the development process, we not only include them in technological advancements but also empower them as influential contributors.

For the elderly, participating in this project is more than just engagement with technology. It presents an opportunity for social interaction, physical activity, and a sense of being valued and heard. Our goal is to create a space where they can express their needs and ideas, thus shaping a robot that truly resonates with their lifestyle and preferences. This initiative is not only about technological development; it is also about improving the quality of life of the elderly through socialization, physical engagement, and the empowerment experience of being an integral part of technological innovation.

Participatory design in the context of people not only facilitates effective application of technology to their needs, but also opens up new possibilities for their decision-making processes. The additional involvement of older people in the design of a social robot provides access to a key technological platform of the future, which is important for their autonomy and dignity. It is also related to social consequences, both with other participants in the project and with the robot itself, which can end up becoming a constant companion. In this way, our initiatives go beyond the boundaries of technology itself, becoming tools to promote social activity, self-fulfillment, and support for older people.

\section{Our PD approach to using social robots for elderly care}

We have four groups of stakeholders participating in our participatory design activities (Fig. 1):
\begin{enumerate}
  \item Elderly Individuals: This is the target user group. Our primary objective is to obtain firsthand perspectives and detailed accounts of their specific needs and daily challenges faced by them
  \item Caregivers: This group includes family, friends, and other non-professional caregivers. They are expected to provide valuable information on the complexities and challenges faced by people who are involved in caring for the elderly on a daily basis.
  \item Medical professionals: This group includes doctors, nurses, therapists, and other professional caregivers. It is important to incorporate their points of view when introducing social robots to mediate interactions with the elderly.
\item Psychologists: We are mentioning this group separately from medical professionals to highlight the role of psychologists who deal with the cognitive and mental health aspects prevalent among the elderly. Their input is expected to shed light on how technological intervention can help with common cognitive impairments and the mental barriers faced by the elderly.
\end{enumerate}

In our approach to participatory design, we embrace the concept of open-ended solutions. Recognizing the rapid pace of technological advancements, we are aware that adapting to these changes can be particularly challenging for the elderly demographic. 

Consequently, our focus is not limited to developing solutions exclusively for the elderly. Instead, our aim is to create versatile tools or systems that can indirectly benefit this group. This involves designing innovations that streamline the process for those who assist the elderly, such as caregivers or healthcare professionals. By easing their efforts, we indirectly improve the quality of life and care of the elderly, acknowledging that the most effective solutions may sometimes be those that support the network surrounding them.

Our user groups also include elderly people with a diverse range of physical and cognitive impairments, such as Alzheimer's and dementia. Consequently, we plan to rely on qualitative analyses in our study. Our goal is to come up with a flexible set of guidelines that can be adapted to people with different needs, and also to different kinds of healthcare ecosystem.

\section{Participatory design activities}

We are implementing our PD approach by conducting the following activities.

\subsection{Identifying the needs of the elderly}
We are using {\it card sorting} and {\it storyboarding} to recognize the needs of the elderly, enabling them to adapt the interface and functions of the robot to their individual requirements (Fig. 1). 
\vspace{\baselineskip}

\textbf{Card sorting}: We are developing a set of cards, based on Cheng et al. \cite{cheng2018essential}, each describing a different function that the robot could potentially perform. These functions range from providing medication reminders, offering companionship, and games that would help them maintain cognitive abilities. After preparing these cards, we will engage each of the four groups of stakeholders mentioned above, inviting them to rank these cards according to what they perceive as the most important. We also provide some blank cards on which they can write their suggestions that are not included in the prepared cards. The insights thus gained from this ranking task are instrumental in determining the design of the robot's functions.

\textbf{Storyboarding:}
We are organizing a diverse group consisting of caregivers, the elderly, and medical professionals for storyboarding sessions. This includes the preparation of storyboard panels with empty cells, each symbolizing a potential step in a scenario. Participants are encouraged to describe various situations where a robot could provide assistance, such as helping with meal preparation or supporting everyday activities. These collaborative storyboarding exercises are expected to yield valuable insights, illustrating how the robot could integrate into the routines of those with neurodegenerative conditions and be tailored to their unique requirements. Through the participation of various stakeholders in the preparation of these storyboards, we can see their points of view.

However, it is crucial to acknowledge some inherent limitations of this method. Storyboards may not cover all possible contexts or situations. Hence, there is a need to be flexible and open to different perspectives. 
\begin{figure}
\centering
\includegraphics[width=0.4\textwidth]{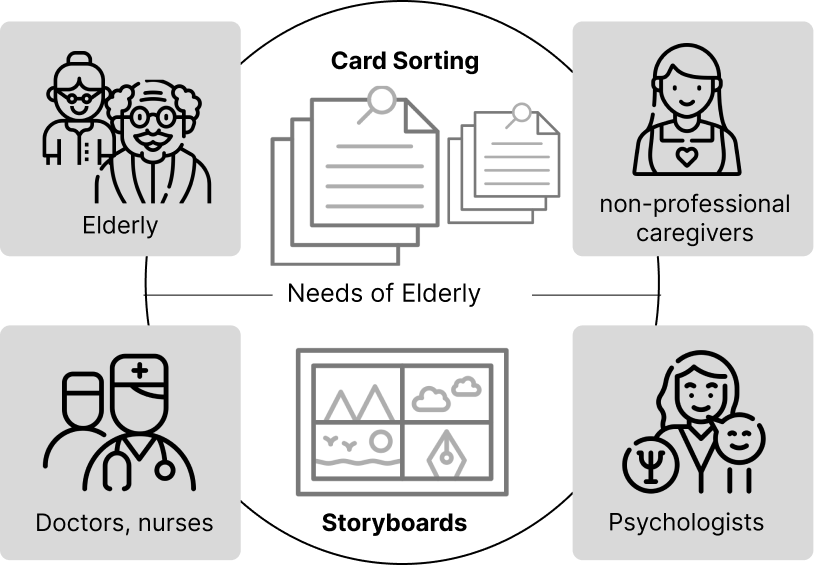}
\caption{Four co-designing Groups in Card Sorting and Storyboards}
\label{fig: Four co-designing groups}
\end{figure}

\subsection{Gaining Multifaceted Perspectives: Semi-Structured Interviews} 

A significant advantage of semi-structured interviews lies in offering participants the opportunity to share their feelings, thoughts, and intimate use case ideas, which they might feel hesitant to express in a group environment \cite{rogers2022maximizing}. Additionally, the semi-structured format offers flexibility and minimal constraints, making it particularly beneficial for activities involving the elderly. One-on-one meetings allow for adjustments in interview length and speaker volume, and is not physically demanding. We are carrying out semi-structured interviews with representatives from each relevant group. The objective is to pose a set of questions designed to explore the following issues.
\begin{enumerate}
  \item Identification of tasks where a robotic assistance could be most advantageous for the elderly. For example, with which tasks do you think you could gain benefits from the robots?
    \item Determination of the most challenging aspects encountered in the daily lives of the elderly. For example, what is the most problematic aspect you face in your daily life?
    \item Insights into what aspects or activities contribute to the happiness and well-being of the elderly population. For example, what makes you happy?
\end{enumerate}

\subsection{Workshop: Engaging the Elderly and Technological Perspectives}

Building on our participatory design approach, we introduce a structured workshop methodology that consists of two distinct parts focused on understanding the elderly and their interactions with technology. In particular, we do not ask directly how to help them, which may induce mixed feelings, especially the feeling of being a burden.

\textbf{Part 1: Elderly Focus: Brainstorming for Empathy and Understanding}

Similarly to interviews, the first segment of the workshop emphasizes understanding the daily challenges of the elderly. Sometimes a group setting has the advantage that an issue that a participant may feel shy to bring up in an individual interview (thinking that the problem is peculiar to them) is easier to discuss in a group when it is realized that others face similar problems. The use of {\it personas}  \cite{adlin2010essential}, a common paradigm for User Experience design, allows participants to project their feelings and concerns on this imaginary visual prop so that they can discuss the issues in an impersonal way \cite{neate2019co}.

Fig. 2 presents the workshop's three stages in part 1 alongside potential answers, clarifying the process, and making it easy to follow. The details of the workshop are as follows:
\begin{itemize}
    \item \textit{Persona Creation}: Participants create personas representing different types of elderly individuals. Each persona will be noted on a separate sticky note, all in one color. 
    \item \textit{Task and Habit Identification}: For each persona, multiple tasks and habits are identified, each depicted on a different sticky note. This visualization technique aims to create {\it flowers} of activities surrounding each persona.
    \item \textit{Jobs-to-be-Done Analysis}: Participants articulate the daily challenges (Jobs-to-be-Done) that these personas face during the tasks and habits identified earlier, focusing on those that are particularly difficult and need a third person. Each Job-to-be-Done is written on a new sticky note.
\end{itemize}
At the end there is an open group analysis, investigating common patterns in the seniors' activities and the most common Jobs-to-be-done.
\begin{figure}
\centering
\includegraphics[width=0.5\textwidth]{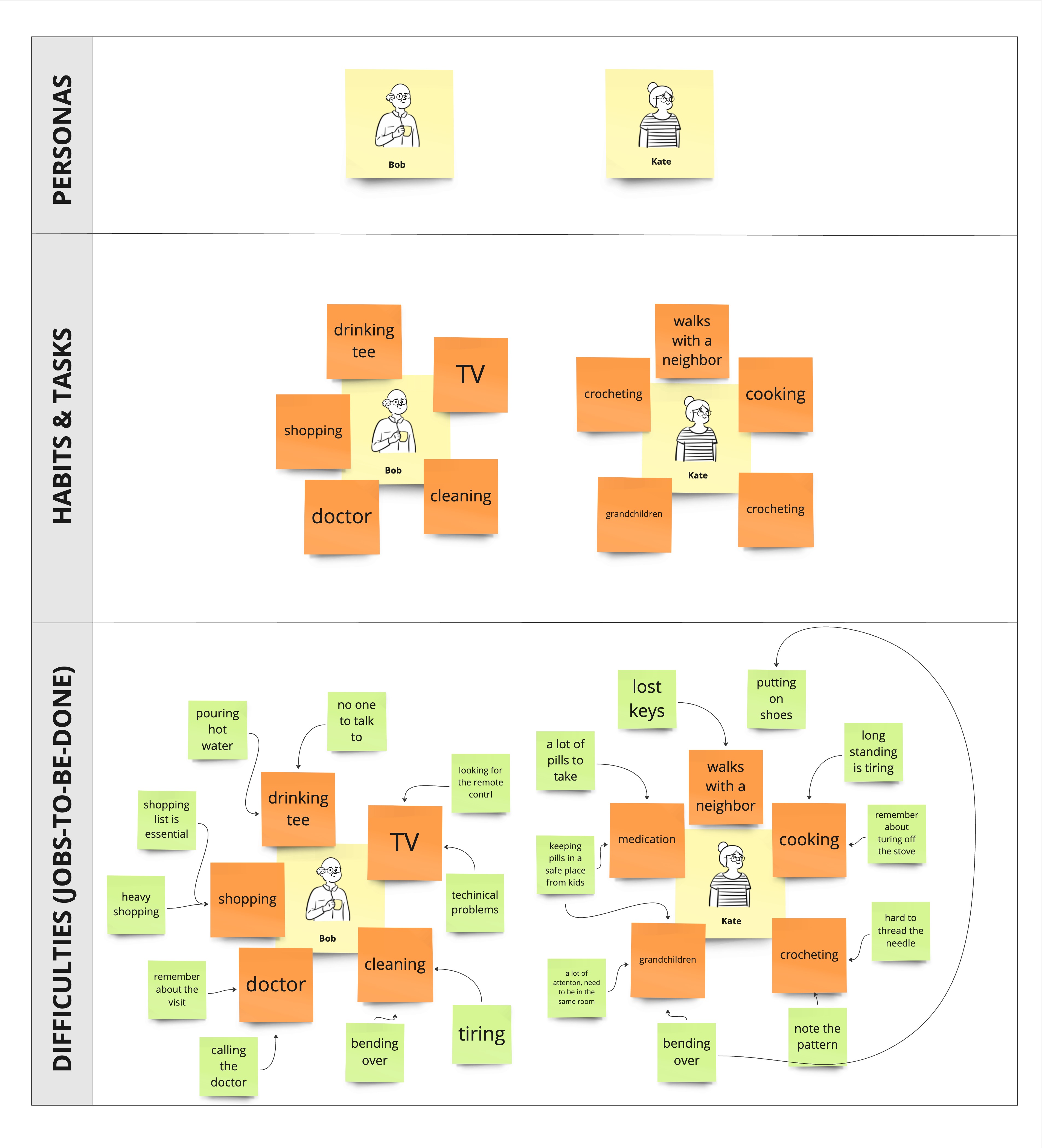}
\caption{Examples of the results of each stage of the workshop}
\label{fig: Examples of the results of each stage of the workshop}
\end{figure}

\textbf{Part 2: Technology Focus - Understanding Technological Challenges}

The second part of the workshop is designed to identify and address the difficulties the elderly  face with technology:
\begin{itemize}
    \item \textit{Identifying Pain Points}: Discussing moments of frustration or helplessness when using technology.
    \item \textit{Problem-Solving Discussions}: Exploring how the elderly manage these technological challenges with the aim of identifying common strategies and solutions.
\end{itemize}

Our approach incorporates two stages (Figs. 3 and 4). In the {\it divergent stage,} activities, such as brainstorming, are carried out to generate a variety of ideas and concepts. In the {\it convergent stage,} these ideas are refined and evaluated, as exemplified by our analysis of Jobs-to-be-Done. This structured methodology ensures a balance between creative exploration and focused development of solutions \cite{liu2003towards}.

\begin{figure} [h!]
\centering
\includegraphics[width=0.35\textwidth]{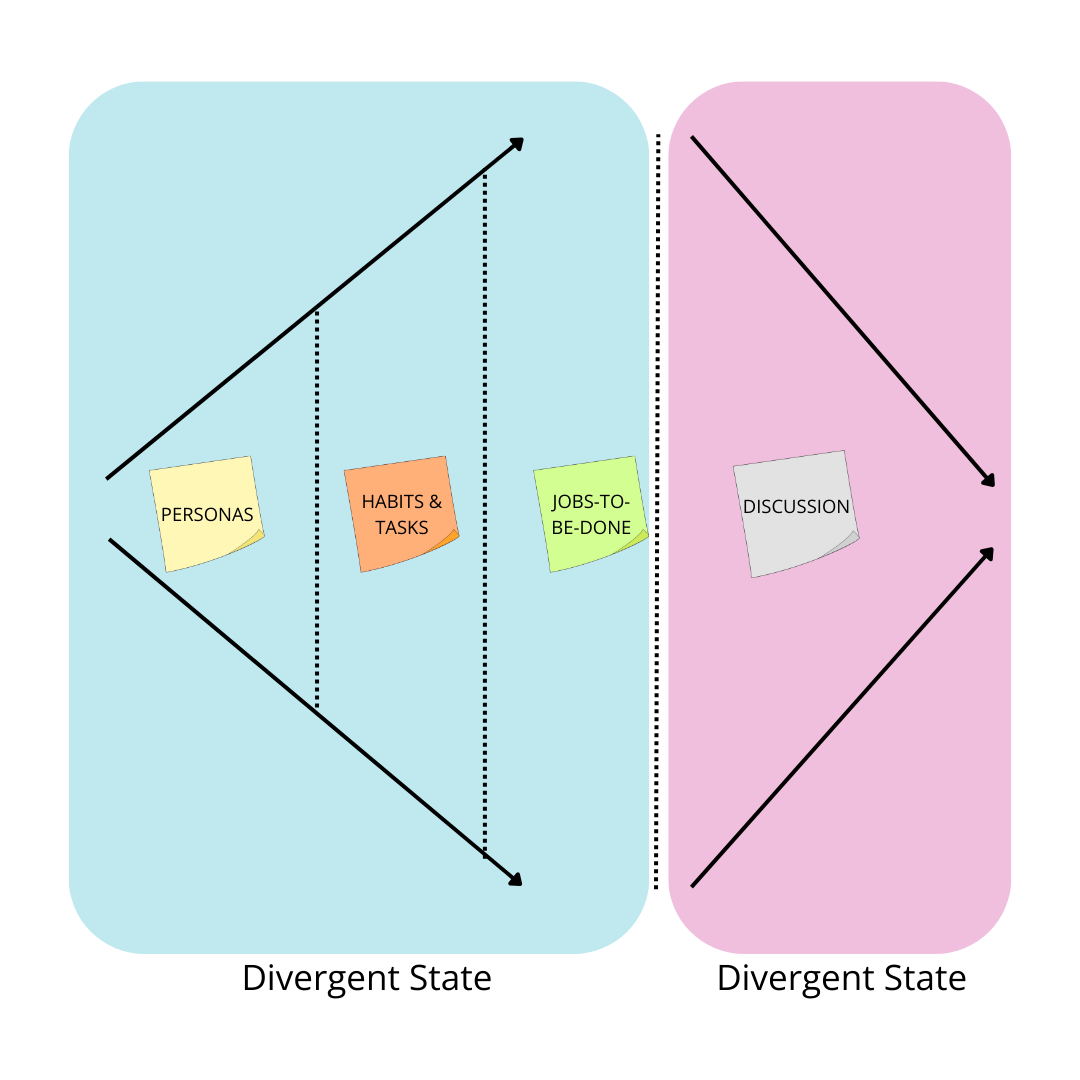}
\caption{Divergent \& Convergent Stages for Part 1: Elderly Focus }
\label{fig:Divergent_Convergent_Stages_for_Part_1_Elderly_Focus}
\end{figure}

\begin{figure} [h!]
\centering
\includegraphics[width=0.35\textwidth]{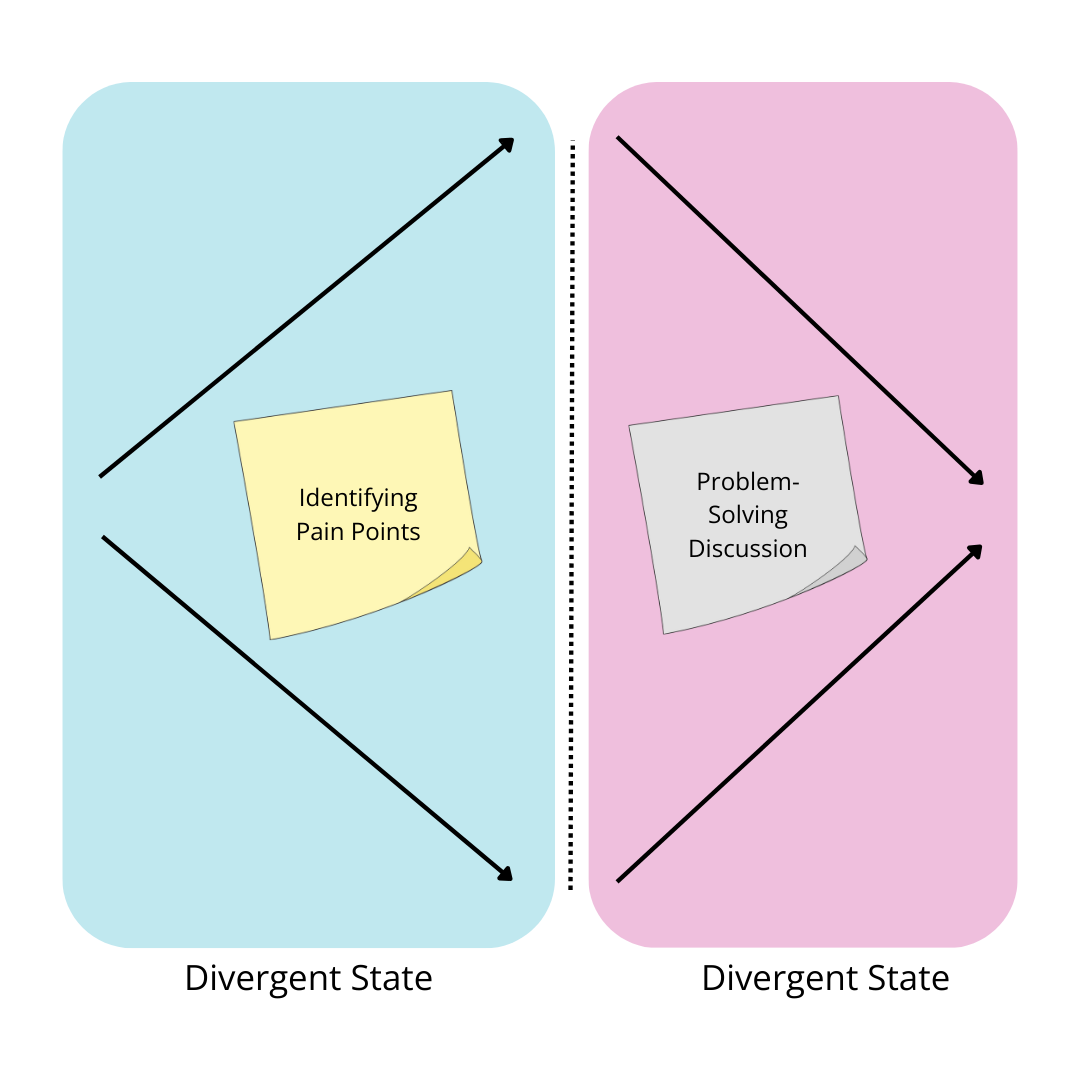}
\caption{Divergent \& Convergent Stages for Part 2: Technology Focus }
\label{fig:Divergent_Convergent_Stages_for_Part_2_Technology_Focus}

\end{figure}


\section{Conclusions}

We presented here our participatory design approach to understand how to effectively deploy social robots for elderly care. Engaging four groups of stakeholders — the elderly, their non-professional caregivers, medical professionals, and psychologists — our research aims to understand and address the needs, concerns, and attitudes towards social robots in care settings. Through a combination of card sorting, storyboarding, and semi-structured interviews, we delve into the perspectives of each stakeholder group. Additionally, our two-phase workshop focuses on identifying both the challenges faced by elderly people in their daily lives and their interactions with technology, with the aim of creating solutions that are not only effective but also empathetic and empowering.


This is our ongoing research, and in future publications we will report on the results of our workshops.




\bibliographystyle{splncs04}
\bibliography{references.bib}

\end{document}